\documentclass[manuscript,screen]{acmart}

\AtBeginDocument{%
  }

\setcopyright{acmlicensed}
\copyrightyear{2018}
\acmYear{2018}
\acmDOI{XXXXXXX.XXXXXXX}

\acmConference[Conference acronym 'XX]{Make sure to enter the correct
  conference title from your rights confirmation emai}{June 03--05,
  2018}{Woodstock, NY}
\acmISBN{978-1-4503-XXXX-X/18/06}

\begin{document}

\title{Boosting Architectural Generation via Prompts: Report}

\author{Xin Zhang}
\author{Wenwen Liu}
\affiliation{%
  \institution{HUST}
  \country{CN}
}


\renewcommand{\shortauthors}{Zhang et al.}


\begin{abstract}

In the realm of AI architectural design, the importance of prompts is becoming increasingly prominent. With advancements in artificial intelligence and large-scale model technology, more design tasks are being delegated to machine learning algorithms. This necessitates a method for designers to guide algorithms in producing their desired designs. Prompts serve as a guiding and motivational mechanism, playing a crucial role in AI-generated architectural design. This paper categorizes and summarizes common vocabulary used in architectural design, discussing how to craft effective prompts and their impact on the quality and creativity of generated results. Through careful prompt design, designers can better control the generated architectural design images, thereby achieving designs that are more aligned with requirements and innovative.
\end{abstract}




\maketitle

\section{Introduction}
In the current field of architectural design \cite{mcguire1983theory,groat2013architectural,garlan1994exploiting,machairas2014algorithms,demirbacs2003focus,caetano2020computational,xu2017taxonomy,kotnik2010digital,hollberg2016lca,aliakseyeu2006computer,akin1996frames,lomas2007architectural}, the development of artificial intelligence 2D/3D technology \cite{Rae2021ScalingLM,Xu2022GroupViTSS,li2023towards,Thoppilan2022LaMDALM, Brown2020LanguageMA, Liu2022FewShotPF} has brought designers both new challenges and opportunities. With the increasing maturity of large-scale model technology \cite{rombach2022stable,openai2023gpt4,Radford2021LearningTV,croitoru2023diffusion}, the application of machine learning algorithms \cite{he2016deep,vaswani2017attention,he2017mask} in generating architectural images is becoming more widespread \cite{li2023sketch,Nauata2020HouseGANRG,li2024generating, chaillou2020archigan,Para2021SketchGenGC}. However, a significant challenge for designers in this process is how to effectively guide algorithms to generate the designs they envision. In this regard, Prompts \cite{jia2022visual,zhou2022learning,liu2022design,li2022align,hao2024optimizing} serve as a crucial mechanism for guiding and motivating.

Prompts are not merely simple words or phrases; they act as communication bridges between designers and machine learning algorithms. Well-designed Prompts can accurately convey design intent and requirements, guiding algorithms to generate architectural images that align with design objectives. In this process, Prompt design needs to consider multiple aspects such as spatial perception, material selection, and functional layout to ensure that the generated designs meet client needs while also demonstrating innovation and uniqueness. Moreover, Prompts can serve as sources of creative inspiration during the design process, stimulating algorithms to generate more imaginative and innovative design solutions.

When using Prompts to guide architectural image generation, designers need to possess deep professional knowledge and extensive design experience. Only by fully understanding the complexity and diversity of architectural design can designers effectively design Prompts that guide algorithms to generate designs that meet requirements. Therefore, Prompt design is not just about arranging words; it also reflects architectural design concepts and aesthetic principles. Designers need to continuously explore and optimize Prompt design methods to better harness the potential of artificial intelligence in architectural design, injecting more novel and unique elements into designs.

This article provides an in-depth summary and classification of commonly used vocabulary in architectural design, aiming to explore how to design Prompts that effectively guide the generation process and analyze their impact on the quality and creativity of generated results.

\section{Architectural Prompts}
The categorized architectural prompts summarize key elements of the architectural design process, spanning from visual perspectives to technical details, and from design concepts to creators. "Perspectives and viewpoints" focus on the observation angles, prompting consideration of how to present buildings and how viewers interact with them. "Lighting Control" centers on utilizing light to shape space and emotional experiences. "Rendering Effects" concentrates on using rendering techniques to showcase the appearance and atmosphere of architectural designs. "Architectural Types" encompass classifications and types of buildings, including residential, commercial, cultural, etc. "Design Styles" cover different design aesthetics, such as modernism, classicism, etc., impacting the appearance and ambiance of buildings. "Building Materials" discusses the selection and application of materials, crucial for both the structure and appearance of buildings. "Architectural Landscapes" focuses on the relationship between buildings and their surrounding environments, including landscape design and urban planning. Lastly, "Architects" highlights the individual design styles of architects; their concepts and creativity shape the appearance and significance of different buildings.

\subsection{Perspectives and viewpoints}
\textit{Three Point Perspective, Aerial Shot, Close-Up, Long Shot, Panoramic View, Dutch Angle, Overhead Shot, High Angle, Low Angle, Point of View (POV), Fish-Eye View, Macro View, Microscopic View, Wide Shot, Full Shot, Medium Shot, Extreme Close-Up, Tilt Shot, Crane Shot, Handheld Shot, Static Shot, Tracking Shot, Zoom Shot, Split Screen, Mirror Shot, POV Shot, Reverse Shot, Establishing Shot, Cutaway Shot, Two Shot, One Shot, Insert Shot, Reaction Shot, Subjective Shot, Objective Shot, Top-Down View, Bottom-Up View, Slanted Angle, Diagonal View, Symmetrical View, Asymmetrical View, Side View, Frontal View.}

\subsection{Lighting Control}
\textit{Bright sunlight, Soft glow, Harsh light, Dim illumination, Radiant beams, Faint gleam, Vivid brightness, Subtle radiance, Intense luminosity, Muted hues, Shining rays, Glimmering light, Sparkling brilliance, Twinkling luminescence, Gleaming brightness, Dazzling brilliance, Blinding glare, Flickering candlelight, Dappled sunlight, Glistening dewdrops, Scattered light, Illuminated room, Fluorescent lighting, Incandescent bulbs, Ambient glow, Transparent sheen, Opaque shadows, Translucent glow, Blurred vision, Clear brightness, Prismatic colors, Refracted light, Reflective surfaces, Penetrating rays, Piercing brightness, Drenched in light, Saturated colors, Patchy illumination, Spotty brightness, Streaked light, Glinting reflections, Glaring sunlight, Searing heat, Illuminated landscape, Radiated warmth, Beaming sunlight, Radiating heat, Glancing reflections, Streaming sunlight, Cascading beams, Sizzling heat, Sputtering flames, Glowing embers, Lit room, Unlit space, Darkened corners, Mellow glow, Muted tones, Reflected light, Diffused illumination, Sharp contrast, Crisp brightness, Precise spotlight, Focused beams, Blurred vision, Distorted shadows, Focused illumination, Vibrant colors, Alight with sunshine, Sunlit room, Moonlit night, Starlit sky, Torchlit path, Candlelit dinner, Streetlit evening, Firelit ambiance, Daylit scene, Nightlit cityscape, Shaded nook, Overcast sky, Sun-drenched landscape, Sun-soaked atmosphere, Light-filled room, Light-splashed surface, Light-dappled forest, Light-streaked sky, Light-flooded space, Light-infused atmosphere, Light-suffused ambiance, Light-imbued surroundings, Light-bathed scenery, Light-gilded landscape.}

\subsection{Rendering Effects}
\textit{Modern, sleek, minimalist, vibrant, dynamic, futuristic, elegant, sophisticated, spacious, airy, luxurious, contemporary, stylish, innovative, geometric, iconic, striking, majestic, picturesque, serene, inviting, seamless, luminous, illuminated, tranquil, polished, refined, detailed, textured, immersive, captivating, artistic, dramatic, panoramic, expansive, panoramic, urban, organic, naturalistic, harmonious, balanced, integrated, sustainable, eco-friendly, technologically advanced, sleek lines, clean design, seamless integration, open concept, fluidity, monumental, grandiose, awe-inspiring, monumental, grandiose, awe-inspiring, ornate, historic, classical, timeless, traditional, cultural, eclectic, dynamic lighting, dramatic shadows, reflective surfaces, transparent elements, panoramic views, focal points, landmark, iconic skyline, futuristic skyline, high-rise, skyscraper, landmark, iconic, cutting-edge, avant-garde, sculptural, abstract, industrial, adaptive reuse, mixed-use, community-oriented, inclusive, accessible, user-friendly, high-tech, energy-efficient, green spaces, outdoor integration, sustainable materials, renewable energy, smart technology, interactive elements, community-centric, collaborative spaces, multipurpose areas, flexible layouts, ergonomic design, seamless transitions, iconic silhouette, dynamic composition, bold colors, soft hues, subtle contrasts, natural materials, harmonious palette, immersive experience, inviting atmosphere.}

\subsection{Architectural Types}
\textit{Hospital, school, university, library, museum, gallery, office building, apartment complex, residential building, condominium, townhouse, villa, mansion, cottage, farmhouse, bungalow, duplex, triplex, skyscraper, high-rise, tower block, shopping mall, retail store, supermarket, department store, boutique, hotel, motel, hostel, inn, guesthouse, lodge, resort, spa, restaurant, cafe, diner, bar, nightclub, theater, cinema, concert hall, stadium, arena, gymnasium, sports complex, swimming pool, recreation center, community center, church, mosque, temple, synagogue, chapel, cathedral, monastery, shrine, mausoleum, cemetery, crematorium, crematory, funeral home, warehouse, factory, manufacturing plant, workshop, garage, hangar, barn, greenhouse, conservatory, laboratory, research facility, observatory, planetarium, clinic, medical center, nursing home, assisted living facility, rehabilitation center, orphanage, shelter, halfway house, prison, jail, courthouse, police station, fire station, embassy, consulate, government building, town hall, city hall, municipal building, post office, bank, financial institution, insurance company, stock exchange, brokerage firm, data center, telecommunications center, power plant, energy facility, water treatment plant, sewage treatment plant, recycling center, landfill, waste management facility, military base, barracks, training facility, command center, missile silo, airfield, airport, seaport, dockyard, lighthouse, observatory, weather station, radio station, television station, broadcasting studio, satellite facility, telecommunications tower. Art studio, art gallery, exhibition hall, performance venue, concert hall, music studio, recording studio, rehearsal space, dance studio, ballet studio, theater, opera house, playhouse, amphitheater, auditorium, lecture hall, conference center, convention center, meeting space, seminar room, classroom, laboratory, research center, innovation hub, coworking space, incubator, accelerator, startup hub, maker space, hackerspace, library, archives, reading room, study hall, digital library, e-library, e-learning center, tutoring center, language school, vocational school, trade school, technical college, community college, liberal arts college, business school, law school, medical school, engineering school, architecture school, design school, art school, conservatory, cooking school, culinary institute, hospitality school, boarding school, day care center, preschool, kindergarten, elementary school, middle school, high school, boarding school, military academy, sports academy, athletic training center, sports complex, stadium, arena, gymnasium, fitness center, yoga studio, martial arts dojo, wellness center, rehabilitation center, physical therapy clinic, medical clinic, health center, dental clinic, eye clinic, chiropractic clinic, holistic health center, veterinary clinic, pet hospital, animal shelter, wildlife sanctuary, botanical garden, arboretum, nature reserve, wildlife preserve, zoo, aquarium, petting zoo, farm, ranch, vineyard, winery, brewery, distillery, tasting room, wine cellar, cellar door, wine bar, cocktail lounge, brewpub, taproom, distillery bar, speakeasy, distillery tasting room, distillery bar, gastropub, bistro, brasserie, trattoria, pizzeria, sushi bar, noodle bar, food court, food truck park, street food market, farmers market, artisanal market, flea market, craft fair, antique market, vintage market, pop-up shop, boutique hotel, boutique inn, boutique resort.}

\subsection{Design Styles}

\textit{Art Deco, Minimalist, Brutalist, Gothic, Baroque, Neoclassical, Contemporary, Victorian, Renaissance, Modernist, Islamic, Romanesque, Tudor, Bauhaus, Industrial, Eclectic, Mediterranean, Prairie Style, Colonial, Federalist, Art Nouveau, Mid-century Modern, Postmodern, International Style, Deconstructivist, Craftsman, Shingle Style, High-Tech, Organic, Futurist, Expressionist, Streamline Moderne, Neo-Gothic, Neo-Renaissance, Neo-Futurist, Neo-Victorian, Neo-Georgian, Neo-Colonial, Neo-Baroque, Neo-Byzantine, Neo-Moorish, Neo-Rococo, Neo-Palladian, Neo-Classical Revival, Neo-Tudor, Neo-Romanesque, Neo-Medieval, Neo-Contemporary, Neo-Industrial, Neo-Traditional, Neo-Modern, Neo-Art Deco, Neo-Eclectic, Neo-Cubist, Neo-Constructivist, Neo-Expressionist, Neo-Primitivist, Neo-Surrealist, Neo-Rocaille, Neo-Enlightenment, Neo-Structuralist, Neo-Functional, Neo-Folk, Neo-Rural, Neo-Suburban, Neo-Urban, Neo-Cosmopolitan, Neo-Nomadic.}

\subsection{Building Materials}
\textit{Concrete, Glass, Steel, Brick, Wood, Stone, Aluminum, Copper, Ceramic, Marble, Granite, Terracotta, Timber, Plastic, Bamboo, Composite materials, Plaster, Fabric, Corrugated metal, Limestone, Slate, Stucco, Rubber, Asphalt, Fiberglass, Plexiglass, Cor-ten steel, Zinc, Wrought iron, Titanium, Polycarbonate, Carbon fiber, Glass fiber, Polyurethane foam, Gypsum, Resin, Clay, Straw, Adobe, Rammed earth, Stainless steel, Copper alloy, Brass, Bronze, Recycled materials, Terrazzo, Vinyl, Polyvinyl chloride (PVC), Acrylic, Engineered wood, Particle board, Medium-density fiberboard (MDF), Oriented strand board (OSB), Laminate, Leather, Cork, Linoleum, Rattan, Bamboo, Hempcrete, Reclaimed wood, Salvaged materials, Gabion, Photovoltaic panels, Solar tiles, Green roof, Living wall, Reflective glass, Insulated concrete forms (ICF), Reinforced concrete, Prefabricated panels, Composite panels, Synthetic turf, Rubberized asphalt, Self-healing concrete, Transparent concrete, Corrugated plastic, Structural insulated panels (SIPs), Translucent polycarbonate, Fiberglass reinforced panels (FRP), Glulam, Cross-laminated timber (CLT), Metal mesh, Expanded metal mesh, Metal cladding, Glass block, Cob, Rammed earth, Cellulose insulation, Aerogel, Structural steel, Engineered stone, Fibre-reinforced polymer (FRP), Hemp fiber, Recycled steel, Bamboo composite, Cast iron, Fiber cement, Poured-in-place concrete, Rubber roofing, Natural fibers, Aluminum composite panel (ACP), Gypsum board, Galvanized steel, Reflective roofing, Composite shingles, Polyethylene, Ceramic tile, Cellular concrete.}

\subsection{Architectural Landscapes}
\textit{
Park, Garden, Plaza, Courtyard, Waterfront, Promenade, Esplanade, Boardwalk, Canal, Riverwalk, Water feature, Fountain, Pond, Lake, Stream, Waterfall, Bridge, Overpass, Underpass, Viaduct, Arch, Tunnel, Pathway, Trail, Walkway, Alley, Avenue, Boulevard, Street, Sidewalk, Promontory, Outlook, Observation deck, Terrace, Patio, Deck, Balcony, Veranda, Porch, Roof garden, Rooftop terrace, Green roof, Sky garden, Arboretum, Botanical garden, Conservatory, Greenhouse, Horticultural display, Woodland, Forest, Grove, Glade, Meadow, Field, Prairie, Pasture, Orchard, Vineyard, Farm, Ranch, Estate, Estate garden, Estate parkland, Nature reserve, Wildlife sanctuary, Wetland, Marsh, Bog, Fen, Swamp, Dune, Desert garden, Rock garden, Alpine garden, Zen garden, Japanese garden, Chinese garden, Mediterranean garden, Tropical garden, Formal garden, Informal garden, Sculpture garden, Labyrinth, Maze, Topiary garden, Parkland, Urban park, City park, National park, State park, Provincial park, Regional park, Community park, Neighborhood park, Memorial park, Sculpture park, Heritage park, Recreational park, Sports park, Adventure park, Water park, Theme park, Botanic park, Cultural park, Historical park, Tourist park, Picnic area, Campground, Outdoor theater, Amphitheater, Bandstand, Performance space, Event venue, Picnic shelter, Barbecue area, Playground, Skate park, Dog park.}

\subsection{Architects}
\textit{
Frank Lloyd Wright, Le Corbusier, Zaha Hadid, Antoni Gaudí, Ludwig Mies van der Rohe, I. M. Pei, Renzo Piano, Louis Kahn, Walter Gropius, Oscar Niemeyer, Eero Saarinen, Richard Meier, Philip Johnson, Frank Gehry, Jean Nouvel, Norman Foster, Tadao Ando, Alvar Aalto, Rem Koolhaas, Santiago Calatrava, Bjarke Ingels, Daniel Libeskind, Moshe Safdie, César Pelli, Herzog de Meuron, Robert Venturi, Denise Scott Brown, Charles Correa, Peter Zumthor, Jean Prouvé, Rafael Moneo, Jeanne Gang, SANAA (Kazuyo Sejima and Ryue Nishizawa), Toyo Ito, Thom Mayne, Álvaro Siza Vieira, Richard Rogers, Marcel Breuer, Jørn Utzon, Skidmore, Owings Merrill (SOM), SOMA (Fernando Romero), Bernard Tschumi, Steven Holl, Bjarke Ingels Group (BIG), Richard Neutra, Paul Rudolph, Kengo Kuma, Cesar Manrique, Paul Williams, Fumihiko Maki, Paul Andreu, Arthur Erickson, Kenzo Tange, William Van Alen, Paul Revere Williams, Charles-Édouard Jeanneret-Gris, Marion Mahony Griffin, Charles and Ray Eames, Eileen Gray, Walter Burley Griffin, Carlo Scarpa, Shigeru Ban, Wang Shu, Richard Buckminster Fuller, Gerrit Rietveld, Louis Sullivan, Gordon Bunshaft, Aldo Rossi, Aldo van Eyck, Frei Otto, Peter Eisenman, Jean Michel Wilmotte, David Chipperfield, Richard England, Henry Hobson Richardson, Kazimir Malevich, Eliel Saarinen, Albert Speer, Oscar Wilde, Will Alsop, Ricardo Bofill, Cedric Price, Dominique Perrault, Ove Arup, Rafael Viñoly, Charles Garnier, Richard Horden, David Adjaye, Winy Maas, Michel Rojkind, Robert Adam, Michael Graves, Tange Kenzō.}

\section{Examples}

\begin{figure*}[tbp]
    \centering
    \includegraphics[width=0.99\textwidth]{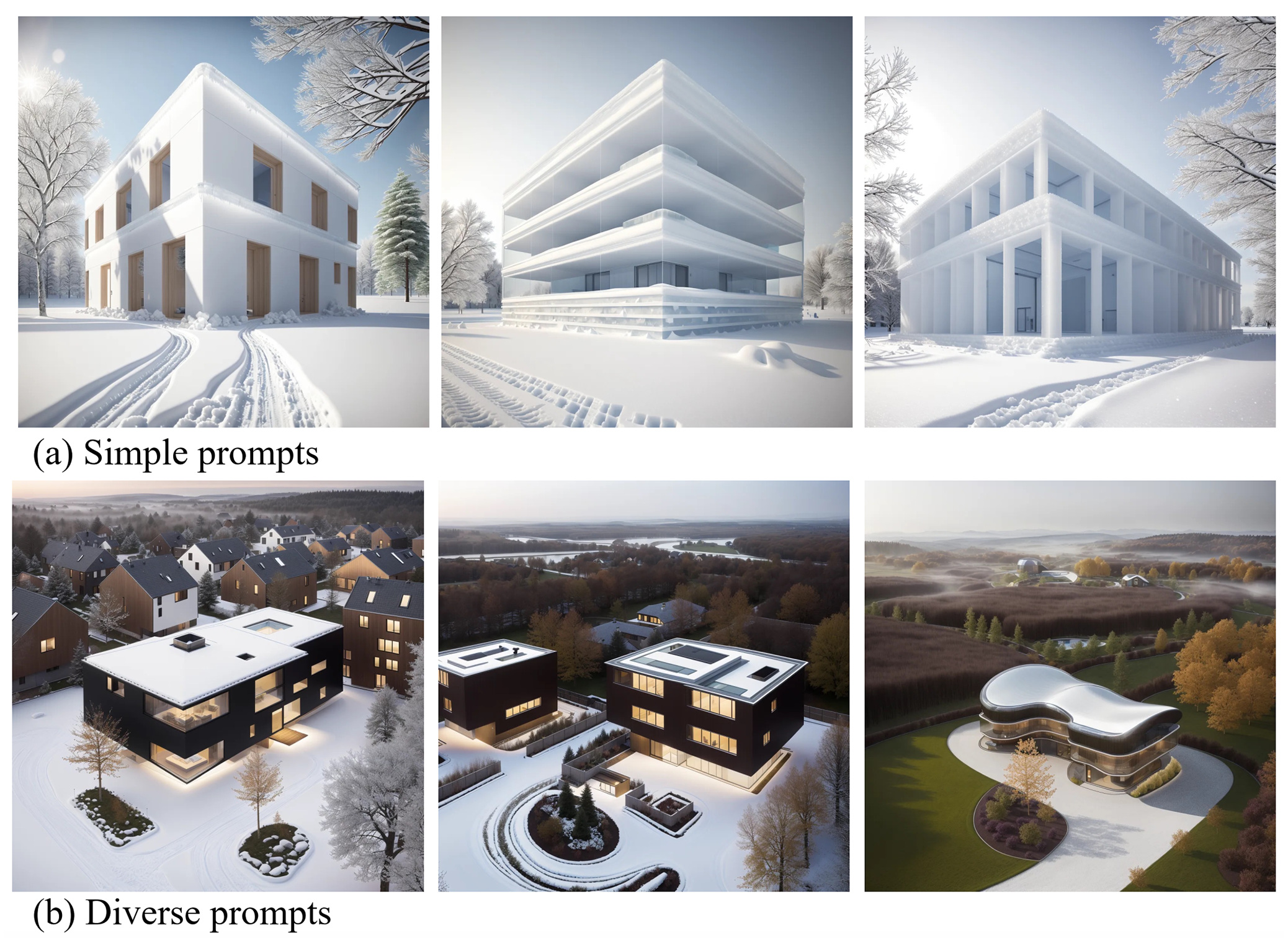}
    \caption{Examples: A Bar. }
\end{figure*} 

\begin{figure*}[tbp]
    \centering
    \includegraphics[width=0.99\textwidth]{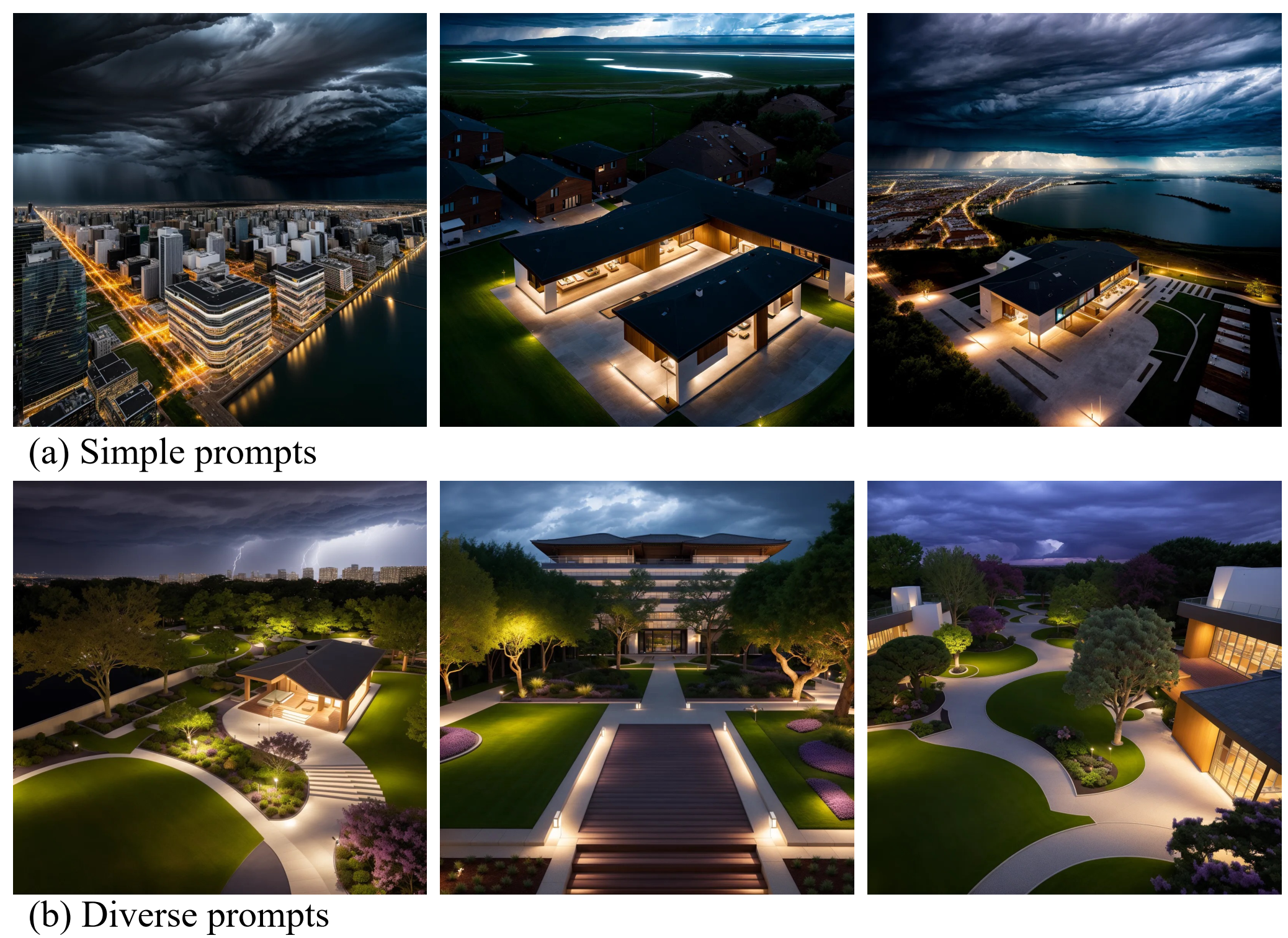}
    \caption{Examples: A Ice and Snow Building. }
\end{figure*} 

In this section, as shown in Fig. 1 and Fig. 2, we demonstrate how finely crafted (a) and diverse prompts (b) can generate more refined architectural images compared to simple prompts.

Prompts of Fig. 1: \textit{
A Bar, Bird's-Eye View, stormy sky illumination, Smooth Shadow Transitions, Bauhaus, terrace, Glass, Japanese Garden, Recreation Park, Waterfront Walkways, Sunflower, Banyan, Wisteria cascades like a waterfall, with strands of purple flowers covering the entire wall, creating a dreamlike scene, Warm, abstract interpretation, Victor Horta.}

Prompts of Fig. 2:\textit{An Ice and Snow Building, Bird's-Eye View, foggy glow, Refined Surface Details, Bauhaus, security camera, Metal, Dutch Garden, Urban Park, Streams, Forsythia, Yew, Grape vines wrap around the walls of a countryside cottage, bearing clusters of translucent grapes in autumn, as if a gift from nature, Morning, high detail, Toyo Ito.}

Prompts 3: \textit{A Department Store, Elevation View, underwater light, Ambient Occlusion, Modernism, columns, Limestone, Japanese Garden, Urban Squares, Canals, Rose, Silk Tree, Boston ivy covers the entire exterior of the building, transforming it into a lively green ecosystem, Rainbow, innovative design, Vincent Callebaut.}

Prompts 4: \textit{A Theater, Perspective, penumbra, Varied Texture Details, Brutalism, wind chimes, Concrete, English Garden, Greenways, Boardwalks, Hydrangea, Cypress, Morning glory spreads along the wall under the morning sun, with its blooming flowers like stars nestled among the green leaves, Snowy, high detail, Harry Seidler.}

\section{Conclusion}

In summary, the advancement of artificial intelligence in architectural design presents both challenges and opportunities. Prompts play a pivotal role in guiding machine learning algorithms to generate designs aligned with designers' visions. Well-crafted Prompts convey design intent effectively, inspiring innovative solutions. However, their success relies on designers' expertise and understanding of architectural complexities. Continuous exploration and refinement of Prompt design methods are essential to fully leverage AI's potential in architectural design. This article provides a thorough exploration of common architectural vocabulary, aiming to empower designers in crafting effective Prompts that enhance the quality and creativity of generated results.

\bibliographystyle{ACM-Reference-Format}
\bibliography{sample-base}

\appendix

\end{document}